\documentclass[aps, prl, reprint,superscriptaddress,citeautoscript]{revtex4-1}
\setcitestyle{super}
\usepackage{graphicx}
\usepackage{epstopdf}
\usepackage{amsmath}

\include{shortcuts}

\begin{document} 

\title{Single quasiparticle trapping in aluminum nanobridge Josephson junctions}
\author{E. M. Levenson-Falk} 
\affiliation{Quantum Nanoelectronics Laboratory, Department of Physics, University of California, Berkeley CA 94720}
\author{F. Kos}
\affiliation{Department of Physics, Yale University, New Haven CT 06520}
\author{R. Vijay}
\affiliation{Quantum Nanoelectronics Laboratory, Department of Physics, University of California, Berkeley CA 94720}
\affiliation{Department of Condensed Matter Physics and Materials Science, Tata Institute of Fundamental Research, Mumbai 400005, India}
\author{L. Glazman} 
\affiliation{Department of Physics, Yale University, New Haven CT 06520}
\author{I. Siddiqi}
\email{irfan\_siddiqi@berkeley.edu}
\affiliation{Quantum Nanoelectronics Laboratory, Department of Physics, University of California, Berkeley CA 94720}

\date{\today}

\begin{abstract}

We present microwave measurements of a high quality factor superconducting resonator incorporating two aluminum nanobridge Josephson junctions in a loop shunted by an on-chip capacitor.  Trapped quasiparticles (QPs) shift the resonant frequency, allowing us to probe the trapped QP number and energy distribution and to quantify their lifetimes.  We find that the trapped QP population obeys a Gibbs distribution above 75 mK, with non-Poissonian trapping statistics. Our results are in quantitative agreement with the Andreev bound state model of transport, and demonstrate a practical means to quantify on-chip QP populations and validate mitigation strategies in a cryogenic environment.
\end{abstract}

%\pacs{42.50.Lc, 42.50.Pq, 03.67.Lx, 85.25.-j}

\maketitle

Superconducting Josephson junctions, with their non-dissipative nonlinearity, form the basis of sensitive magnetometers\cite{SQUIDhandbook,Levenson-Falk2013}, ultra-low-noise analog amplifiers\cite{Hatridge2011,DevoretJPC,KonradJPA}, and quantum bits \cite{qubitreview}.  The presence of quasiparticle (QP) excitations degrades or even disrupts the operation of many of these circuits.  If QPs tunnel across or become trapped in a junction, they cause dissipation or excess noise that can limit qubit coherence and reduce the sensitivity of amplifier and magnetometer devices. Recent measurements have confirmed that such nonequilibrium excitations are present in superconducting circuits, even at very low tempeatures\cite{fluxonium,Paik2011,MartinisHeater}, well below the energy gap.  Moreover, QP tunneling dynamics have also been studied in superconducting qubits\cite{Riste2013}, providing valuable information on QP lifetimes. 

%By probing the behavior of QPs trapped in a junction, it is possible to learn about their distribution and thermalization mechanisms.

In this letter, we quantify the trapping behavior of QPs in a resonant LC circuit consisting of a capacitively shunted nanoscale superconducting quantum interference device (SQUID). In this architecture, the Josephson junctions are not tunnel junctions but rather submicron constrictions with $\approx 700$ conduction channels, each of which can be viewed as having a pair of quantized energy levels with a magnetic-flux-dependent gap. At finite flux bias, the trapping probability is large and a single trapped QP effectively neutralizes one channel, resulting in a shift of the junction kinetic inductance and correspondingly the resonant frequency of the SQUID circuit. The shift due to a \emph{single} trapped QP is readily resolved provided the quality factor ($Q$) of the resonator is sufficiently high. We use this effect to infer the average number of trapped QPs ($\overline{n}_{trap}$) and their energy distribution function. Furthermore, using microwave pulses and time domain measurements we resonantly excite the QPs, infer the gap between the trap states and the quasiparticle continuum ($\Delta_A$), and measure the associated trapping time. Our measurements are in quantitative agreement with the Andreev level picture of conduction channels, and illustrate a powerful method to characterize nonequilibrium QP populations in superconducting devices\textemdash a valuable new tool for optimizing quantum circuits.

In the semiconductor representation of a Josephson junction, the supercurrent is carried by a set of conduction channels, with the $i$th channel having a transmittivity $\tau_i$ between 0 and 1.  For each channel, there is a pair of Andreev bound states with energies
\begin{equation}
\label{eq:EA}
E_{A\pm}^{(i)}(\delta) =\pm \Delta \sqrt{1 - \tau_i \sin^2{\frac{\delta}{2}}}\,,
\end{equation}
where $\delta$ is the gauge-invariant superconducting phase difference across the junction and $\Delta$ is the superconducting gap\cite{Kulik1970,LikharevWeakLink,Beenakker1991}.  Energy spectra for $\tau$ = 1 and 0.5 are shown in Fig. 1(a). At zero temperature, the lower bound state is occupied and the upper state is unoccupied.  However, for $\tau_i > 0$ and $\delta > 0$, the upper bound state energy drops below $+ \Delta$, and so forms a subgap trap state for any QPs which may exist in the bulk superconductor.  The upper bound state, when occupied by a QP, carries current in the opposite direction as the lower state, and so occupation of both states leads to a total critical current of zero for that conduction channel.  This ``poisoning'' of a conduction channel causes a shift in both the critical current and the inductance of the junction, as the poisoned channel no longer participates in supercurrent transport.  Previous work has probed the trapping of QPs in few-channel quantum point contact junctions by performing switching measurements of the critical current\cite{Zgirski2011, Bretheau2012, Bretheau2013}.  In contrast, our approach realizes a dispersive measurement of the Josephson inductance of many-channel nanobridge junctions.

\begin{figure}
\includegraphics{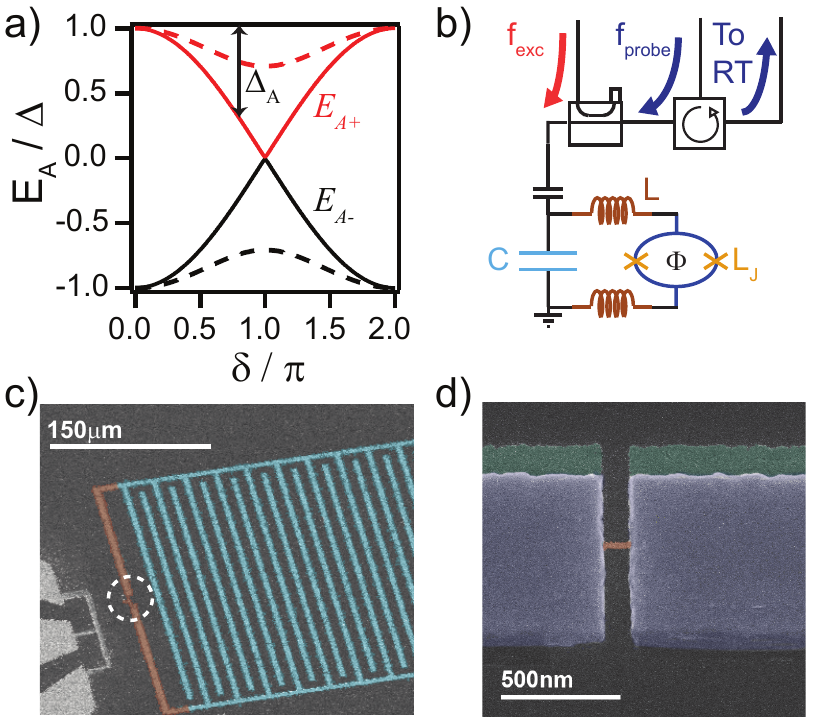} 
\caption{\label{fig1} (Color online) (a) Illustration of the Andreev bound states for channels with $\tau = 1$ (solid lines) and $\tau = 0.5$ (dashed lines).  The Andreev gap $\Delta_A$ is illustrated for the $\tau = 1$ channel. (b) Simplified measurement schematic.  The resonator is biased with a static flux $\Phi_{\mathrm{dc}}$; measurements are performed by passing a weak probe tone (blue) through a circulator and analyzing the reflected signal.  A strong excitation tone (red) may be applied via a directional coupler in order to alter the distribution of QPs in the junctions.  (c) False-color SEM image of device, showing the nanoSQUID (circled), meander inductor (orange), and IDC (blue).  A superconducting fast flux line, not used in this experiment, is shown off to the left in grey. (d) Detailed false-color SEM image of the nanobridge junction, showing the bridge (orange) connecting the thick banks (blue and green, highlighting the two layers of evaporation).}
\end{figure}

Our device consists of two variable-thickness aluminum nanobridge Josephson junctions arranged symmetrically in a small SQUID loop. This geometry allows us to apply a phase bias $\delta =\pi \phi$ to both junctions, where $\phi \equiv \Phi / \Phi_0$ is the normalized flux through the loop in terms of the magnetic flux quantum $\Phi_0$. Fabrication details are contained in the supplementary material.  The nanobridges are 8 nm thick, 100 nm long, and 25 nm wide, connecting banks which are 80 nm thick and 750 nm wide, in a $2\times2 \ \mu$m SQUID loop.  The SQUID is placed in series with a linear inductance of 1.2 nH and this combination is shunted by an interdigitated capacitor (IDC) to form a weakly nonlinear parallel LC oscillator with resonant frequency $\omega_0 = 2\pi \times 4.72$ GHz at zero magnetic flux. SEM images of the device are shown in Fig. 1(c) and (d). The oscillator is isolated from the external environment by small IDCs, causing it to be critically coupled with $Q_{ext} \approx Q_{int} \approx 2Q = 5.3 \times 10^4$, thus giving a linewidth of 180 kHz.  A simplified schematic of our measurement setup is shown in Fig. 1(b).  The oscillator is enclosed in multiple layers of thermal and magnetic shielding, and attached to the base stage of a cryogen-free dilution refrigerator with a base temperature of 10 mK.  Static flux bias is applied via a superconducting coil which sits underneath the small copper box housing the device.  We perform microwave reflectometry, injecting power via heavily attenuated lines through a cryogenic circulator and passing the reflected signal through a semiconductor HEMT amplifier at 3 K before multiple amplification stages at room temperature.  The sample temperature may be accurately controlled using a resistive heater at the base stage.

\begin{figure}[tbp]
\includegraphics{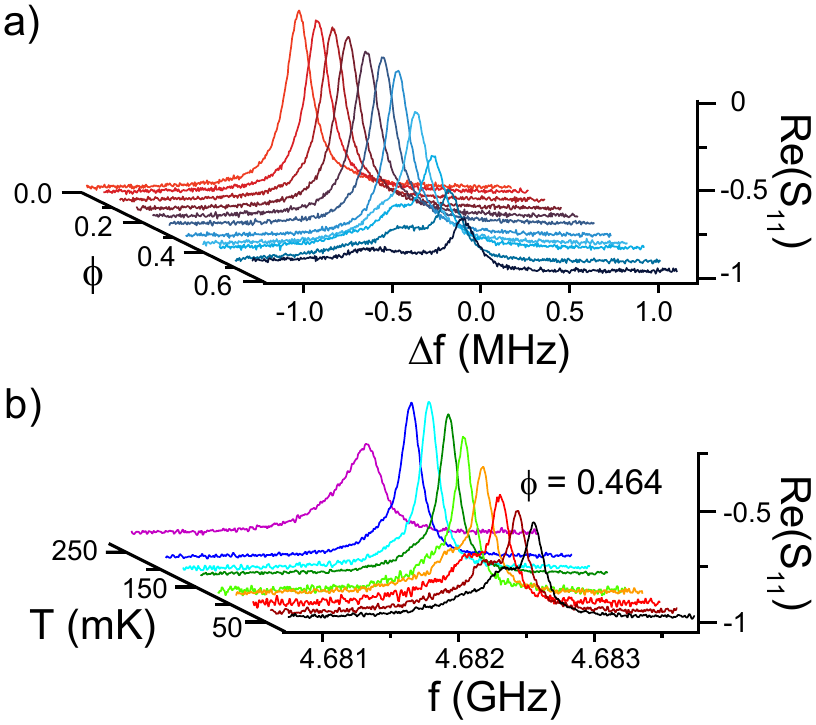} 
\caption{\label{fig2} (Color online) (a) Resonance lineshapes at different flux biases, plotted as a function of frequency detuning from the 0-QP resonance.  The single Lorentzian peak present at 0 flux begins to develop multiple broad humps as flux bias increases. (b) Temperature dependence of the resonance lineshape, showing the decrease in QP trapping as temperature rises.  All data are shown at $\phi = 0.464$.  At the lowest temperature, multiple humps'are resolvable, indicating multiple QP trapping numbers.  At higher temperatures, first the 2-QP and then the 1-QP peak become suppressed, leading to a single Lorentzian resonance.  At even higher temperatures (T = 250 mK), the resonance broadens, shrinks, and moves to lower frequency; we attribute this to loss originating from bulk QP transport in the resonator, as the QP density becomes quite high.}
\end{figure}

Resonance curves as a function of flux, plotted as the real part of the reflected signal, are shown in Fig. 2(a).  At low flux bias ($\phi \lesssim 0.2$), the resonance has an Lorentzian lineshape.  However, at larger flux biases the resonance peak shrinks and begins to develop a second broad hump at lower frequency, indicating the development of another resonance peak; at the highest flux biases the resonance shows a second, even broader hump (resembling a long tail) at even lower frequency.  This behavior is indicative of QPs trapping in the junction, raising its inductance and thus lowering the resonant frequency of the device.  By averaging many resonance traces together on a timescale that is long compared to the QP dynamics (see supplementary material), we integrate over all possible configurations of QPs in the junction, weighted by their probabilities.  Thus, the resonance lineshape tells us the average configuration of QPs trapped in the junction.  As flux bias grows, $E_A$ drops, and so the likelihood of a QP trapping in the Andreev excited state grows.  This means that the probability that 0 QPs are trapped shrinks, and so the associated 0-QP resonance peak shrinks.  In a junction with many conduction channels with different values of $\tau$ there are a range of values of $E_A(\tau)$, leading to differing trapping probabilities.  Since the different channels have different untrapped inductances, there is a distribution of resonant frequencies with 1 trapped QP $\omega_0(\tau)$, resulting in a broad 1-QP resonance peak.  A similar argument applies to the 2-QP peak, where the range of trapping configurations is even broader.

We next measure the resonance at finite flux bias as a function of temperature; data at $\phi = 0.464$ is shown in Fig. 2(b).  At low temperatures, 2 broad humps are resolvable in the resonance curves, indicating trapping of 1 and 2 QPs.  As temperature rises within the range 10-200 mK, first the 2-QP and then the 1-QP hump are suppressed, as $\overline{n}_{trap}$ decreases.  This is to be expected, as hotter QPs are less likely to occupy lower-energy states (i.e. trap states).  At higher temperatures (T $= 200$ mK), $\overline{n}_{trap}$ starts to rise again due to increasing QP density in the resonator.  At even higher temperatures (T $= 250$ mK), the resonance peak becomes shorter and broader while shifting to lower frequency, even at 0 flux.  We attribute this effect to increased loss in the resonator due to bulk QP transport, as the normalized quasiparticle density $x_{qp}$ becomes quite large at 250 mK.  

\begin{figure}
\includegraphics{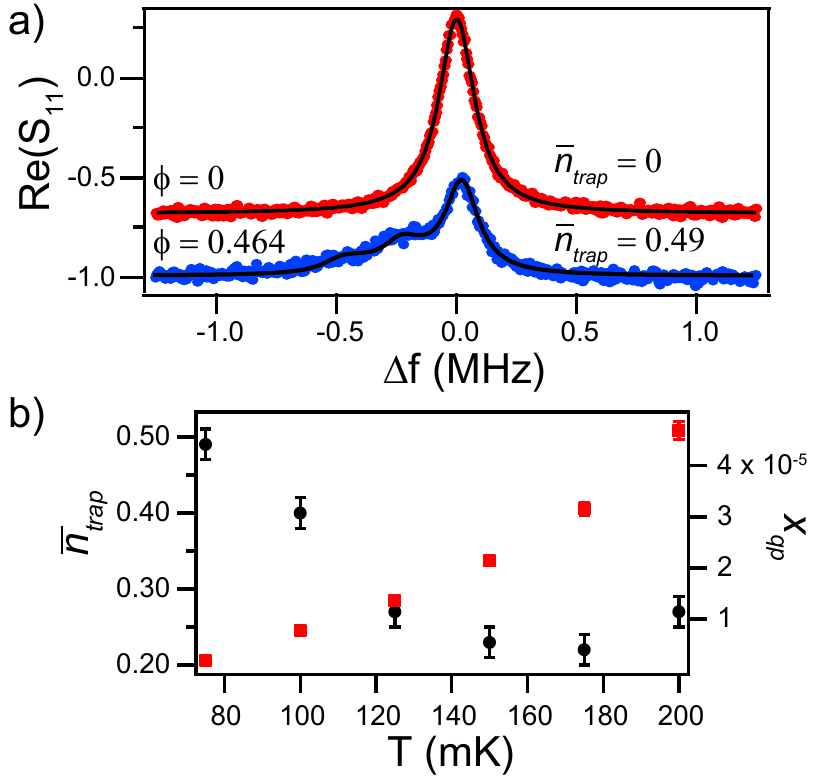} 
\caption{\label{fig3} (Color online) (a) Frequency response at 100 mK for $\phi = 0$ (top, red) and 0.464 (bottom, blue), illustrating the change in resonance lineshape.  Traces are vertically offset for clarity.  The black curves are fits using Eq.~(\ref{eq:SLong}), showing excellent agreement with the data. (b) $\overline{n}_{trap}$ (left axis, black circles) at $\phi$ = 0.464 and extracted values of $x_{qp}$ (right axis, red squares) as a function of temperature.}
\end{figure}

We next attempt to quantitatively describe the QP trapping behavior by fitting the resonance curves.  We first extract the junction inductance $L_J(\phi)$ by fitting the flux tuning curve of the resonance to find the participation ratio $q(\phi)$ of the junction inductance in the total inductance (see supplementary material).  This procedure gives a value  $L_J(0) \approx 35$ pH for each junction.  Using the Dorokhov distribution \cite{Dorokhov1982} of the channel transmittivities $\rho(\tau)$ (applicable to diffusive nanobridges), we find the effective number of channels $N_e = 3\hbar^2 / (2\Delta e^2  L_J) = 683$.  We define the QP configuration $\{n_i\}$, where $n_i = 0,1$ is the number of QPs in the $i$th channel, neglecting the possibility of having two QPs in the same channel since the number of channels is large.  The resonant frequency $\omega_0$ and thus the response function $S(\omega)$ of the oscillator is a function of this configuration, $S(\omega,\omega_0(\{n_i\}))$.  We measure the response on time scales which are long compared to the QP dynamics, and thus average over all configurations weighted by their probabilities:
\begin{equation}
 \overline{S}(\omega) = \sum_i p(\{n_i\}) S(\omega,\omega_0(\{n_i\}))\,.
\end{equation}
We consider only the resonances resulting from 0-2 trapped QPs, weighted by their probabilities $P_{0,1,2}$, with $\overline{n}_{trap} = P_1 + 2 P_2$.  Thus, the resonator response is given by
\begin{widetext}
\begin{equation}
\label{eq:SLong}
\overline{S}(\omega) = P_0 S(\omega, \omega_0^{(0)}) + P_1\int\limits_0^1d\tau\rho(\tau)f(E_A(\tau))S(\omega,\omega_0(\tau))+  P_2\int\limits_0^1d\tau_1\int\limits_0^1d\tau_2 \ \rho(\tau_1)\rho(\tau_2)f(E_A(\tau_1))f(E_A(\tau_2))S(\omega,\omega_0(\tau_1,\tau_2))\,.
\end{equation}
\end{widetext}
Here $\omega_0^{(0)}$, $\omega_0(\tau)$, and $\omega(\tau_1,\tau_2)$ are the resonant frequencies with no trapped QPs, a single QP trapped in Andreev level formed by channel of transmittivity $\tau$, and two QPs trapped in two Andreev states (channels characterized by $\tau_1$ and $\tau_2$), respectively. We assume that the trapped QPs obey a Gibbs distribution, $f(E_A)\propto e^{-E_A/k_BT}$.  We fit the resonance with the $P_k$ as the only free parameters; sample fits at T = 75 mK for $\phi =0$ and 0.464 are shown in Fig 3(a).  The theory produces excellent agreement with the data for $\overline{n}_{trap} \lesssim 1$, suggesting that trapped QPs follow a thermal Gibbs distribution with temperature equal to the fridge temperature of 75 mK.  We note that in regimes where $\overline{n}_{trap} \sim 1$, the 3-QP contribution to the resonance is non-negligible at the few percent level; this causes the fit $P_k$ to sum to less than 1.  Fitting the 3-QP resonance is computationally intensive, and so we restrict our analysis to the first two QP peaks.  We note that in general the $P_k$ are not Poisson-distributed (even accounting for the neglected 3-QP peak). \footnote{For $T=100$ mK the measured values are $P_0=0.66$, $P_1=0.16$, $P_2=0.12$, while the Poisson distribution $P_k=\bar{n}_{trap}^k \exp(-\bar{n}_{trap}) /k!$ gives $P_0=0.66$, $P_1=0.26$, $P_2=0.06$. }

We repeat our fitting procedure at all temperatures below 250 mK. Values of $\overline{n}_{trap}$ between 75-200 mK are plotted in Fig. 3(b), as well as extracted values of $x_{qp}$.  The 75 mK value of $x_{qp} = 1 \times 10^{-6}$ is consistent with other measurements of aluminum superconducting circuits \cite{Paik2011,MartinisQPs}.  This indicates that our method can reproduce the results of other techniques for measuring the mean density of QPs, while also shedding light on their distribution.  Although our theory fits the data well in the range 75-200 mK, we find that below 75 mK we cannot fit the data with a Gibbs distribution at the fridge temperature.  There are likely two causes for this discrepancy: at low temperatures $\overline{n}_{trap}$ grows, and so the 3-QP peak becomes significant; also, below 75 mK the QPs may not be thermally distributed, as thermalization mechanisms such as inelastic electron-phonon scattering will be suppressed at low temperatures due to the falling phonon density\cite{ShumeikoPhonon}.  The overall population of nonequilibrium QPs is likely due to remnant infrared radiation leaking through our sample shielding and impacting the device.  Indeed, a systematic reduction in $x_{qp}$ was observed with the addition of several layers of radiation shielding over the course of these experiments; the reported data correspond to the lowest value of $x_{qp}$ observed.  This highlights the utility of this simple circuit as a tool for characterizing excess radiation leakage. 

Our QP trapping theory can also be applied in a different limit, where $\overline{n}_{trap}$ is large enough such that $P_0 \approx 0$.  In this limit, one may use the saddle-point approximation to integrate over a continuous quasiparticle occupation (rather than summing discrete occupation numbers). We can then write the resonant response as a Lorentzian centered at $\omega_0(\overline{n}_{trap}) = \omega_0^{(0)} + \sum_i p_i \frac{\partial \omega_0}{\partial n_i}$, convolved with a Gaussian of width $\delta\omega_0^2 = \sum_i p_i (\frac{\partial \omega_0}{\partial n_i})^2$, where $p_i$ is the probability of having a QP in $i$th channel.  To reach this regime, we deliberately raise $x_{qp}$ and thus $\overline{n}_{trap}$ by heating a radiator near the sample box.  Infrared radiation leaks into the box and hits the resonator, creating QPs \cite{MartinisHeater}.  Fig.~4 shows a resonance curve at 100 mK and $\phi = 0.432$, with a theory fit (dashed line) giving $\overline{n}_{trap} = 7$.  Also plotted for comparison is a resonance at zero flux, which is well-fit by a Lorentzian (indicating $\overline{n}_{trap} = 0$).  The best fit to the $\phi=0.432$ data (solid line) is obtained by allowing the Gaussian width to vary independently from the Lorentzian centerpoint; extracting QP number from each gives $\overline{n}_{trap} = 19$ and 7, respectively.  The discrepancy is likely due to the limitations of the short-channel approximation accepted in Eq.~(\ref{eq:EA})\cite{Vijay2009,Vijay2010}; it may also be related to the non-Poissonian nature of the low-$\overline{n}_{trap}$ data, since the Gaussian is just the high-$\overline{n}_{trap}$ limit of Poisson statistics.

\begin{figure}[tbp]
\includegraphics{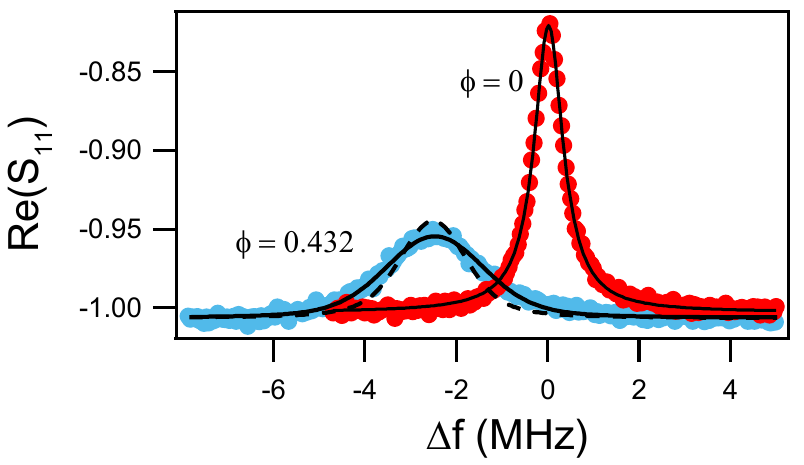} 
\caption{\label{fig4} (Color online) Resonance traces taken with the radiator on, showing a Lorentzian lineshape  at 0 flux (right, red) and a Lorentzian  convolved with a Gaussian (with a center frequency which is lower than its 0-QP value) at $\phi = 0.432$ (left, blue).  Fits using the high-$\overline{n}_{trap}$ Gaussian approximation are shown in black. The dashed black line is a fit constrained by a single parameter $\overline{n}_{trap} = 7$, while the solid lines allow the width and centerpoint of the convoluted Gaussian-Lorentzian to vary independently (giving $\overline{n}_{trap} = 19$ and 7, respectively).  The zero-flux data is well-fit by a Lorentzian, indicating $\overline{n}_{trap} = 0$.}
\end{figure}

By illuminating the junction with a microwave tone of frequency $f_{exc} > \Delta_A / h \equiv (\Delta - E_A(\delta, \tau)) / h$, it is possible to promote a QP out of its Andreev trap and back up into the continuum, thus restoring the relevant conduction channel to its original properties (see Fig. 1(a)).  When such an excitation tone is on, the resonant response retains its Lorentzian lineshape at all flux biases, with only a small decrease in the 0-QP resonance peak at high flux bias.  We attribute this small decrease to an imperfect excitation efficiency, as the QP trapping probability becomes larger at large phase biases, but the microwave excitation power (i.e. QP excitation rate) is limited, as an excitation tone at very high power will begin to bias the junctions into nonlinearity and shift the resonance.  The fact that the resonance lineshape remains constant as a function of flux indicates that when no QPs are trapped other effects such as flux noise and phase slips in the junctions are negligible.

In order to probe the energy spectrum of trapped QPs, we sweep the frequency of the applied microwave excitation tone and measure the response of the resonator as a function of flux (i.e. phase bias).  Spectroscopy data at 100 mK is shown in Fig.~5, plotted as the real part of the reflected probe signal at the 0-QP resonant frequency minus its value with the bias tone off; a positive shift indicates a greater probability of 0 QPs, and thus shows that trapped QPs have been cleared from the junction.  At a given flux bias, a minimum excitation tone frequency is required to observe any response; this frequency grows with increasing flux bias, and has a value consistent with $\Delta_A(\phi,\tau)$ for $\tau \approx 0.8-1$.  At high excitation frequencies ($f_{exc} > \Delta_A(\phi, 1)$), the response saturates.  The width of this transition from no response to saturation is consistent with the Dorokhov distribution for channel transmittivities.  Finally, we note that the saturated response grows with flux bias, as the 0-QP resonance with no bias becomes more suppressed and thus the effect of the excitation tone becomes more pronounced.

\begin{figure}[tbh]
\includegraphics{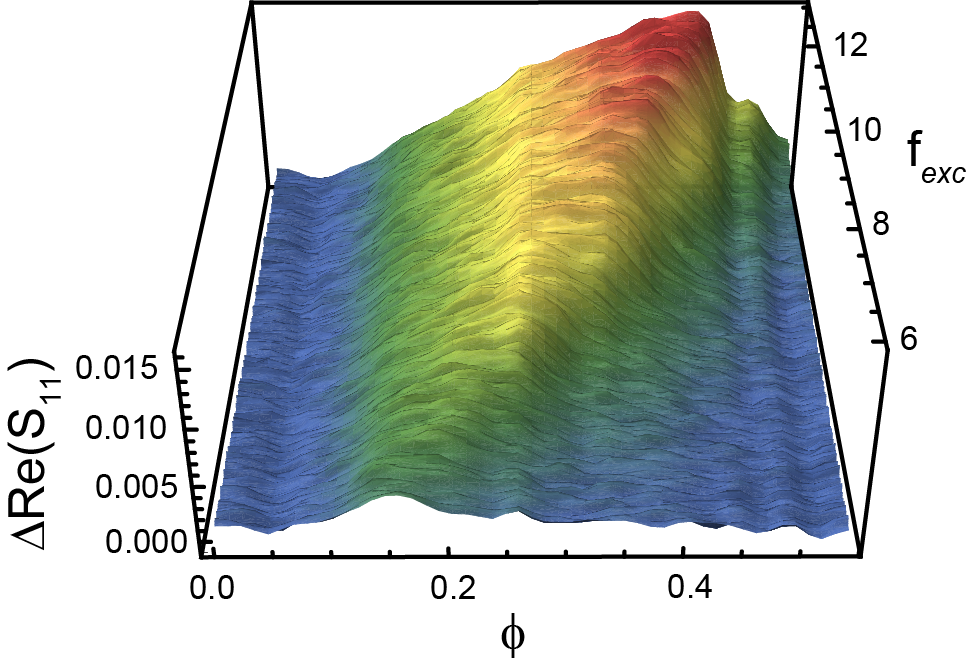} 
\caption{\label{fig5} (Color online) Spectroscopy data, showing the shift in the 0-QP resonance peak height versus flux $\phi$ and excitation tone frequency $f_{exc}$, referred to its height with the excitation tone off.  The data show a threshhold excitation frequency which grows with flux, consistent with $\Delta_A(\phi)$, and a height shift which grows with flux, indicating a growing $\overline{n}_{trap}$. }
\end{figure}

By pulsing the excitation tone and monitoring the response of the 0-QP resonance, it is possible to probe the dynamics associated with QP trapping and excitation.  Data at base temperature are shown in the supplementary material; during both trapping and excitation the resonance evolves exponentially in time towards steady state.  The excitation time scales inversely with excitation tone power and is relatively constant as a function of flux.  Typical values range from 1 to 5 $\mu$s for the excitation powers used.  The retrapping time shortens as flux increases, ranging from 60 to 15 $\mu$s.  The scaling as a function of flux appears consistent with electron-phonon relaxation being the dominant mechanism for QP trapping, although the basic theory does not correctly predict the order of magnitude of the retrapping time (see supplementary material for more details).

In conclusion, we have measured ensemble QP trapping in a phase-biased nanobridge.  By using a dispersive measurement of a narrow-linewidth resonator, we are able to resolve single QPs trapped in junctions with $\sim 700$ channels, while keeping the junction in the superconducting state at fixed phase bias.  The trapped QPs obey a thermal Gibbs distribution for temperatures above 75 mK, both in the limit of low and high average trapped QP number.  Applying a microwave bias tone to the junction results in a frequency-dependent clearing of trapped QPs, with a retrapping time ranging from 15-60 $\mu$s.  Future work can futher probe the mechanisms of QP trapping and thermalization and investigate any correlations between trapping/untrapping events in nearby junctions.  By engineering the parameters of the resonant circuit, it should be possible to achieve continuous single-shot measurement of the QP configuration.  Finally, we note that the number of trapped QPs is a sensitive probe of the QP density, thus allowing our device to be used to evaluate the quality of the radiation shielding used in a cryogenic setup.

\begin{acknowledgements}
The authors acknowledge N. Roch, {\c{C.}} Girit, G. Catelani, M. Devoret, A. Eddins, S. Hacohen-Gourgy, and R. Schoelkopf for useful discussions. Funding was provided in part by an NSF GRFP grant (EMLF) at UC Berkeley, by AFOSR under Grant FA9550-08-1-0104, by DOE contract DE-FG02-08ER46482 (theory of resonance line shapes), and by IARPA via the Army Research Office W911NF-09-1-0369 (analysis of experimental data) at Yale University.
\end{acknowledgements}

\bibliography{QPtrapping}

%merlin.mbs apsrev4-1.bst 2010-07-25 4.21a (PWD, AO, DPC) hacked
%Control: key (0)
%Control: author (8) initials jnrlst
%Control: editor formatted (1) identically to author
%Control: production of article title (-1) disabled
%Control: page (0) single
%Control: year (1) truncated
%Control: production of eprint (0) enabled
\begin{thebibliography}{22}%
\makeatletter
\providecommand \@ifxundefined [1]{%
 \@ifx{#1\undefined}
}%
\providecommand \@ifnum [1]{%
 \ifnum #1\expandafter \@firstoftwo
 \else \expandafter \@secondoftwo
 \fi
}%
\providecommand \@ifx [1]{%
 \ifx #1\expandafter \@firstoftwo
 \else \expandafter \@secondoftwo
 \fi
}%
\providecommand \natexlab [1]{#1}%
\providecommand \enquote  [1]{``#1''}%
\providecommand \bibnamefont  [1]{#1}%
\providecommand \bibfnamefont [1]{#1}%
\providecommand \citenamefont [1]{#1}%
\providecommand \href@noop [0]{\@secondoftwo}%
\providecommand \href [0]{\begingroup \@sanitize@url \@href}%
\providecommand \@href[1]{\@@startlink{#1}\@@href}%
\providecommand \@@href[1]{\endgroup#1\@@endlink}%
\providecommand \@sanitize@url [0]{\catcode `\\12\catcode `\$12\catcode
  `\&12\catcode `\#12\catcode `\^12\catcode `\_12\catcode `\%12\relax}%
\providecommand \@@startlink[1]{}%
\providecommand \@@endlink[0]{}%
\providecommand \url  [0]{\begingroup\@sanitize@url \@url }%
\providecommand \@url [1]{\endgroup\@href {#1}{\urlprefix }}%
\providecommand \urlprefix  [0]{URL }%
\providecommand \Eprint [0]{\href }%
\providecommand \doibase [0]{http://dx.doi.org/}%
\providecommand \selectlanguage [0]{\@gobble}%
\providecommand \bibinfo  [0]{\@secondoftwo}%
\providecommand \bibfield  [0]{\@secondoftwo}%
\providecommand \translation [1]{[#1]}%
\providecommand \BibitemOpen [0]{}%
\providecommand \bibitemStop [0]{}%
\providecommand \bibitemNoStop [0]{.\EOS\space}%
\providecommand \EOS [0]{\spacefactor3000\relax}%
\providecommand \BibitemShut  [1]{\csname bibitem#1\endcsname}%
\let\auto@bib@innerbib\@empty
%</preamble>
\bibitem [{\citenamefont {Clarke}\ and\ \citenamefont
  {Braginski}(2004)}]{SQUIDhandbook}%
  \BibitemOpen
  \bibfield  {author} {\bibinfo {author} {\bibfnamefont {J.}~\bibnamefont
  {Clarke}}\ and\ \bibinfo {author} {\bibfnamefont {A.~I.}\ \bibnamefont
  {Braginski}},\ }\href@noop {} {\emph {\bibinfo {title} {{The SQUID
  Handbook}}}}\ (\bibinfo  {publisher} {Wiley Online Library},\ \bibinfo {year}
  {2004})\BibitemShut {NoStop}%
\bibitem [{\citenamefont {Levenson-Falk}\ \emph {et~al.}(2013)\citenamefont
  {Levenson-Falk}, \citenamefont {Vijay}, \citenamefont {Antler},\ and\
  \citenamefont {Siddiqi}}]{Levenson-Falk2013}%
  \BibitemOpen
  \bibfield  {author} {\bibinfo {author} {\bibfnamefont {E.~M.}\ \bibnamefont
  {Levenson-Falk}}, \bibinfo {author} {\bibfnamefont {R.}~\bibnamefont
  {Vijay}}, \bibinfo {author} {\bibfnamefont {N.}~\bibnamefont {Antler}}, \
  and\ \bibinfo {author} {\bibfnamefont {I.}~\bibnamefont {Siddiqi}},\ }\href
  {\doibase 10.1088/0953-2048/26/5/055015} {\bibfield  {journal} {\bibinfo
  {journal} {Superconductor Science and Technology}\ }\textbf {\bibinfo
  {volume} {26}},\ \bibinfo {pages} {055015} (\bibinfo {year}
  {2013})}\BibitemShut {NoStop}%
\bibitem [{\citenamefont {Hatridge}\ \emph {et~al.}(2011)\citenamefont
  {Hatridge}, \citenamefont {Vijay}, \citenamefont {Slichter}, \citenamefont
  {Clarke},\ and\ \citenamefont {Siddiqi}}]{Hatridge2011}%
  \BibitemOpen
  \bibfield  {author} {\bibinfo {author} {\bibfnamefont {M.}~\bibnamefont
  {Hatridge}}, \bibinfo {author} {\bibfnamefont {R.}~\bibnamefont {Vijay}},
  \bibinfo {author} {\bibfnamefont {D.~H.}\ \bibnamefont {Slichter}}, \bibinfo
  {author} {\bibfnamefont {J.}~\bibnamefont {Clarke}}, \ and\ \bibinfo {author}
  {\bibfnamefont {I.}~\bibnamefont {Siddiqi}},\ }\href {\doibase
  10.1103/PhysRevB.83.134501} {\bibfield  {journal} {\bibinfo  {journal} {Phys.
  Rev. B}\ }\textbf {\bibinfo {volume} {83}},\ \bibinfo {pages} {134501}
  (\bibinfo {year} {2011})}\BibitemShut {NoStop}%
\bibitem [{\citenamefont {Abdo}\ \emph {et~al.}(2011)\citenamefont {Abdo},
  \citenamefont {Schackert}, \citenamefont {Hatridge}, \citenamefont
  {Rigetti},\ and\ \citenamefont {Devoret}}]{DevoretJPC}%
  \BibitemOpen
  \bibfield  {author} {\bibinfo {author} {\bibfnamefont {B.}~\bibnamefont
  {Abdo}}, \bibinfo {author} {\bibfnamefont {F.}~\bibnamefont {Schackert}},
  \bibinfo {author} {\bibfnamefont {M.}~\bibnamefont {Hatridge}}, \bibinfo
  {author} {\bibfnamefont {C.}~\bibnamefont {Rigetti}}, \ and\ \bibinfo
  {author} {\bibfnamefont {M.}~\bibnamefont {Devoret}},\ }\href {\doibase
  10.1063/1.3653473} {\bibfield  {journal} {\bibinfo  {journal} {Applied
  Physics Letters}\ }\textbf {\bibinfo {volume} {99}},\ \bibinfo {eid} {162506}
  (\bibinfo {year} {2011})}\BibitemShut {NoStop}%
\bibitem [{\citenamefont {Castellanos-Beltran}\ \emph
  {et~al.}(2009)\citenamefont {Castellanos-Beltran}, \citenamefont {Irwin},
  \citenamefont {Vale}, \citenamefont {Hilton},\ and\ \citenamefont
  {Lehnert}}]{KonradJPA}%
  \BibitemOpen
  \bibfield  {author} {\bibinfo {author} {\bibfnamefont {M.}~\bibnamefont
  {Castellanos-Beltran}}, \bibinfo {author} {\bibfnamefont {K.}~\bibnamefont
  {Irwin}}, \bibinfo {author} {\bibfnamefont {L.}~\bibnamefont {Vale}},
  \bibinfo {author} {\bibfnamefont {G.}~\bibnamefont {Hilton}}, \ and\ \bibinfo
  {author} {\bibfnamefont {K.}~\bibnamefont {Lehnert}},\ }\href {\doibase
  10.1109/TASC.2009.2018119} {\bibfield  {journal} {\bibinfo  {journal}
  {Applied Superconductivity, IEEE Transactions on}\ }\textbf {\bibinfo
  {volume} {19}},\ \bibinfo {pages} {944} (\bibinfo {year} {2009})}\BibitemShut
  {NoStop}%
\bibitem [{\citenamefont {Devoret}\ and\ \citenamefont
  {Schoelkopf}(2013)}]{qubitreview}%
  \BibitemOpen
  \bibfield  {author} {\bibinfo {author} {\bibfnamefont {M.~H.}\ \bibnamefont
  {Devoret}}\ and\ \bibinfo {author} {\bibfnamefont {R.~J.}\ \bibnamefont
  {Schoelkopf}},\ }\href {\doibase 10.1126/science.1231930} {\bibfield
  {journal} {\bibinfo  {journal} {Science}\ }\textbf {\bibinfo {volume}
  {339}},\ \bibinfo {pages} {1169} (\bibinfo {year} {2013})}\BibitemShut
  {NoStop}%
\bibitem [{\citenamefont {Manucharyan}\ \emph {et~al.}(2009)\citenamefont
  {Manucharyan}, \citenamefont {Koch}, \citenamefont {Glazman},\ and\
  \citenamefont {Devoret}}]{fluxonium}%
  \BibitemOpen
  \bibfield  {author} {\bibinfo {author} {\bibfnamefont {V.~E.}\ \bibnamefont
  {Manucharyan}}, \bibinfo {author} {\bibfnamefont {J.}~\bibnamefont {Koch}},
  \bibinfo {author} {\bibfnamefont {L.~I.}\ \bibnamefont {Glazman}}, \ and\
  \bibinfo {author} {\bibfnamefont {M.~H.}\ \bibnamefont {Devoret}},\ }\href
  {\doibase 10.1126/science.1175552} {\bibfield  {journal} {\bibinfo  {journal}
  {Science}\ }\textbf {\bibinfo {volume} {326}},\ \bibinfo {pages} {113}
  (\bibinfo {year} {2009})}\BibitemShut {NoStop}%
\bibitem [{\citenamefont {Paik}\ \emph {et~al.}(2011)\citenamefont {Paik},
  \citenamefont {Schuster}, \citenamefont {Bishop}, \citenamefont {Kirchmair},
  \citenamefont {Catelani}, \citenamefont {Sears}, \citenamefont {Johnson},
  \citenamefont {Reagor}, \citenamefont {Frunzio}, \citenamefont {Glazman},
  \citenamefont {Girvin}, \citenamefont {Devoret},\ and\ \citenamefont
  {Schoelkopf}}]{Paik2011}%
  \BibitemOpen
  \bibfield  {author} {\bibinfo {author} {\bibfnamefont {H.}~\bibnamefont
  {Paik}}, \bibinfo {author} {\bibfnamefont {D.~I.}\ \bibnamefont {Schuster}},
  \bibinfo {author} {\bibfnamefont {L.~S.}\ \bibnamefont {Bishop}}, \bibinfo
  {author} {\bibfnamefont {G.}~\bibnamefont {Kirchmair}}, \bibinfo {author}
  {\bibfnamefont {G.}~\bibnamefont {Catelani}}, \bibinfo {author}
  {\bibfnamefont {A.~P.}\ \bibnamefont {Sears}}, \bibinfo {author}
  {\bibfnamefont {B.~R.}\ \bibnamefont {Johnson}}, \bibinfo {author}
  {\bibfnamefont {M.~J.}\ \bibnamefont {Reagor}}, \bibinfo {author}
  {\bibfnamefont {L.}~\bibnamefont {Frunzio}}, \bibinfo {author} {\bibfnamefont
  {L.~I.}\ \bibnamefont {Glazman}}, \bibinfo {author} {\bibfnamefont {S.~M.}\
  \bibnamefont {Girvin}}, \bibinfo {author} {\bibfnamefont {M.~H.}\
  \bibnamefont {Devoret}}, \ and\ \bibinfo {author} {\bibfnamefont {R.~J.}\
  \bibnamefont {Schoelkopf}},\ }\href {\doibase 10.1103/PhysRevLett.107.240501}
  {\bibfield  {journal} {\bibinfo  {journal} {Phys. Rev. Lett.}\ }\textbf
  {\bibinfo {volume} {107}},\ \bibinfo {pages} {240501} (\bibinfo {year}
  {2011})}\BibitemShut {NoStop}%
\bibitem [{\citenamefont {Barends}\ \emph {et~al.}(2011)\citenamefont
  {Barends}, \citenamefont {Wenner}, \citenamefont {Lenander}, \citenamefont
  {Chen}, \citenamefont {Bialczak}, \citenamefont {Kelly}, \citenamefont
  {Lucero}, \citenamefont {O'Malley}, \citenamefont {Mariantoni}, \citenamefont
  {Sank}, \citenamefont {Wang}, \citenamefont {White}, \citenamefont {Yin},
  \citenamefont {Zhao}, \citenamefont {Cleland}, \citenamefont {Martinis},\
  and\ \citenamefont {Baselmans}}]{MartinisHeater}%
  \BibitemOpen
  \bibfield  {author} {\bibinfo {author} {\bibfnamefont {R.}~\bibnamefont
  {Barends}}, \bibinfo {author} {\bibfnamefont {J.}~\bibnamefont {Wenner}},
  \bibinfo {author} {\bibfnamefont {M.}~\bibnamefont {Lenander}}, \bibinfo
  {author} {\bibfnamefont {Y.}~\bibnamefont {Chen}}, \bibinfo {author}
  {\bibfnamefont {R.~C.}\ \bibnamefont {Bialczak}}, \bibinfo {author}
  {\bibfnamefont {J.}~\bibnamefont {Kelly}}, \bibinfo {author} {\bibfnamefont
  {E.}~\bibnamefont {Lucero}}, \bibinfo {author} {\bibfnamefont
  {P.}~\bibnamefont {O'Malley}}, \bibinfo {author} {\bibfnamefont
  {M.}~\bibnamefont {Mariantoni}}, \bibinfo {author} {\bibfnamefont
  {D.}~\bibnamefont {Sank}}, \bibinfo {author} {\bibfnamefont {H.}~\bibnamefont
  {Wang}}, \bibinfo {author} {\bibfnamefont {T.~C.}\ \bibnamefont {White}},
  \bibinfo {author} {\bibfnamefont {Y.}~\bibnamefont {Yin}}, \bibinfo {author}
  {\bibfnamefont {J.}~\bibnamefont {Zhao}}, \bibinfo {author} {\bibfnamefont
  {A.~N.}\ \bibnamefont {Cleland}}, \bibinfo {author} {\bibfnamefont {J.~M.}\
  \bibnamefont {Martinis}}, \ and\ \bibinfo {author} {\bibfnamefont {J.~J.~A.}\
  \bibnamefont {Baselmans}},\ }\href@noop {} {\bibfield  {journal} {\bibinfo
  {journal} {Applied Physics Letters}\ }\textbf {\bibinfo {volume} {99}},\
  \bibinfo {eid} {113507} (\bibinfo {year} {2011})}\BibitemShut {NoStop}%
\bibitem [{\citenamefont {Rist\`{e}}\ \emph {et~al.}(2013)\citenamefont
  {Rist\`{e}}, \citenamefont {Bultink}, \citenamefont {Tiggelman},
  \citenamefont {Schouten}, \citenamefont {Lehnert},\ and\ \citenamefont
  {DiCarlo}}]{Riste2013}%
  \BibitemOpen
  \bibfield  {author} {\bibinfo {author} {\bibfnamefont {D.}~\bibnamefont
  {Rist\`{e}}}, \bibinfo {author} {\bibfnamefont {C.~C.}\ \bibnamefont
  {Bultink}}, \bibinfo {author} {\bibfnamefont {M.~J.}\ \bibnamefont
  {Tiggelman}}, \bibinfo {author} {\bibfnamefont {R.~N.}\ \bibnamefont
  {Schouten}}, \bibinfo {author} {\bibfnamefont {K.~W.}\ \bibnamefont
  {Lehnert}}, \ and\ \bibinfo {author} {\bibfnamefont {L.}~\bibnamefont
  {DiCarlo}},\ }\href {\doibase 10.1038/ncomms2936} {\bibfield  {journal}
  {\bibinfo  {journal} {Nature communications}\ }\textbf {\bibinfo {volume}
  {4}},\ \bibinfo {pages} {1913} (\bibinfo {year} {2013})}\BibitemShut
  {NoStop}%
\bibitem [{\citenamefont {{Kulik}}(1969)}]{Kulik1970}%
  \BibitemOpen
  \bibfield  {author} {\bibinfo {author} {\bibfnamefont {I.~O.}\ \bibnamefont
  {{Kulik}}},\ }\href@noop {} {\bibfield  {journal} {\bibinfo  {journal}
  {Soviet Journal of Experimental and Theoretical Physics}\ }\textbf {\bibinfo
  {volume} {30}},\ \bibinfo {pages} {944} (\bibinfo {year} {1969})}\BibitemShut
  {NoStop}%
\bibitem [{\citenamefont {Likharev}(1979)}]{LikharevWeakLink}%
  \BibitemOpen
  \bibfield  {author} {\bibinfo {author} {\bibfnamefont {K.~K.}\ \bibnamefont
  {Likharev}},\ }\href {\doibase 10.1103/RevModPhys.51.101} {\bibfield
  {journal} {\bibinfo  {journal} {Rev. Mod. Phys.}\ }\textbf {\bibinfo {volume}
  {51}},\ \bibinfo {pages} {101} (\bibinfo {year} {1979})}\BibitemShut
  {NoStop}%
\bibitem [{\citenamefont {Beenakker}\ and\ \citenamefont {van
  Houten}(1991)}]{Beenakker1991}%
  \BibitemOpen
  \bibfield  {author} {\bibinfo {author} {\bibfnamefont {C.~W.~J.}\
  \bibnamefont {Beenakker}}\ and\ \bibinfo {author} {\bibfnamefont
  {H.}~\bibnamefont {van Houten}},\ }\href {\doibase
  10.1103/PhysRevLett.66.3056} {\bibfield  {journal} {\bibinfo  {journal}
  {Phys. Rev. Lett.}\ }\textbf {\bibinfo {volume} {66}},\ \bibinfo {pages}
  {3056} (\bibinfo {year} {1991})}\BibitemShut {NoStop}%
\bibitem [{\citenamefont {Zgirski}\ \emph {et~al.}(2011)\citenamefont
  {Zgirski}, \citenamefont {Bretheau}, \citenamefont {Le~Masne}, \citenamefont
  {Pothier}, \citenamefont {Esteve},\ and\ \citenamefont
  {Urbina}}]{Zgirski2011}%
  \BibitemOpen
  \bibfield  {author} {\bibinfo {author} {\bibfnamefont {M.}~\bibnamefont
  {Zgirski}}, \bibinfo {author} {\bibfnamefont {L.}~\bibnamefont {Bretheau}},
  \bibinfo {author} {\bibfnamefont {Q.}~\bibnamefont {Le~Masne}}, \bibinfo
  {author} {\bibfnamefont {H.}~\bibnamefont {Pothier}}, \bibinfo {author}
  {\bibfnamefont {D.}~\bibnamefont {Esteve}}, \ and\ \bibinfo {author}
  {\bibfnamefont {C.}~\bibnamefont {Urbina}},\ }\href@noop {} {\bibfield
  {journal} {\bibinfo  {journal} {Physical review letters}\ }\textbf {\bibinfo
  {volume} {106}},\ \bibinfo {pages} {257003} (\bibinfo {year}
  {2011})}\BibitemShut {NoStop}%
\bibitem [{\citenamefont {Bretheau}\ \emph {et~al.}(2012)\citenamefont
  {Bretheau}, \citenamefont {Girit}, \citenamefont {Tosi}, \citenamefont
  {Goffman}, \citenamefont {Joyez}, \citenamefont {Pothier}, \citenamefont
  {Esteve},\ and\ \citenamefont {Urbina}}]{Bretheau2012}%
  \BibitemOpen
  \bibfield  {author} {\bibinfo {author} {\bibfnamefont {L.}~\bibnamefont
  {Bretheau}}, \bibinfo {author} {\bibfnamefont {{\c{C}}.}~\bibnamefont
  {Girit}}, \bibinfo {author} {\bibfnamefont {L.}~\bibnamefont {Tosi}},
  \bibinfo {author} {\bibfnamefont {M.}~\bibnamefont {Goffman}}, \bibinfo
  {author} {\bibfnamefont {P.}~\bibnamefont {Joyez}}, \bibinfo {author}
  {\bibfnamefont {H.}~\bibnamefont {Pothier}}, \bibinfo {author} {\bibfnamefont
  {D.}~\bibnamefont {Esteve}}, \ and\ \bibinfo {author} {\bibfnamefont
  {C.}~\bibnamefont {Urbina}},\ }\href@noop {} {\bibfield  {journal} {\bibinfo
  {journal} {Comptes Rendus Physique}\ }\textbf {\bibinfo {volume} {13}},\
  \bibinfo {pages} {89} (\bibinfo {year} {2012})}\BibitemShut {NoStop}%
\bibitem [{\citenamefont {Bretheau}\ \emph {et~al.}(2013)\citenamefont
  {Bretheau}, \citenamefont {Girit}, \citenamefont {Pothier}, \citenamefont
  {Esteve},\ and\ \citenamefont {Urbina}}]{Bretheau2013}%
  \BibitemOpen
  \bibfield  {author} {\bibinfo {author} {\bibfnamefont {L.}~\bibnamefont
  {Bretheau}}, \bibinfo {author} {\bibfnamefont {{\c{C}}.~O.}\ \bibnamefont
  {Girit}}, \bibinfo {author} {\bibfnamefont {H.}~\bibnamefont {Pothier}},
  \bibinfo {author} {\bibfnamefont {D.}~\bibnamefont {Esteve}}, \ and\ \bibinfo
  {author} {\bibfnamefont {C.}~\bibnamefont {Urbina}},\ }\href {\doibase
  10.1038/nature12315} {\bibfield  {journal} {\bibinfo  {journal} {Nature}\
  }\textbf {\bibinfo {volume} {499}},\ \bibinfo {pages} {312} (\bibinfo {year}
  {2013})}\BibitemShut {NoStop}%
\bibitem [{\citenamefont {Dorokhov}(1982)}]{Dorokhov1982}%
  \BibitemOpen
  \bibfield  {author} {\bibinfo {author} {\bibfnamefont {O.}~\bibnamefont
  {Dorokhov}},\ }\href@noop {} {\bibfield  {journal} {\bibinfo  {journal} {JETP
  Lett}\ }\textbf {\bibinfo {volume} {36}},\ \bibinfo {pages} {318} (\bibinfo
  {year} {1982})}\BibitemShut {NoStop}%
\bibitem [{Note1()}]{Note1}%
  \BibitemOpen
  \bibinfo {note} {For $T=100$ mK the measured values are $P_0=0.66$,
  $P_1=0.16$, $P_2=0.12$, while the Poisson distribution $P_k=\protect
  \mathaccentV {bar}016{n}_{trap}^k \protect \qopname \relax o{exp}(-\protect
  \mathaccentV {bar}016{n}_{trap}) /k!$ gives $P_0=0.66$, $P_1=0.26$,
  $P_2=0.06$.}\BibitemShut {Stop}%
\bibitem [{\citenamefont {Lenander}\ \emph {et~al.}(2011)\citenamefont
  {Lenander}, \citenamefont {Wang}, \citenamefont {Bialczak}, \citenamefont
  {Lucero}, \citenamefont {Mariantoni}, \citenamefont {Neeley}, \citenamefont
  {O'Connell}, \citenamefont {Sank}, \citenamefont {Weides}, \citenamefont
  {Wenner}, \citenamefont {Yamamoto}, \citenamefont {Yin}, \citenamefont
  {Zhao}, \citenamefont {Cleland},\ and\ \citenamefont
  {Martinis}}]{MartinisQPs}%
  \BibitemOpen
  \bibfield  {author} {\bibinfo {author} {\bibfnamefont {M.}~\bibnamefont
  {Lenander}}, \bibinfo {author} {\bibfnamefont {H.}~\bibnamefont {Wang}},
  \bibinfo {author} {\bibfnamefont {R.~C.}\ \bibnamefont {Bialczak}}, \bibinfo
  {author} {\bibfnamefont {E.}~\bibnamefont {Lucero}}, \bibinfo {author}
  {\bibfnamefont {M.}~\bibnamefont {Mariantoni}}, \bibinfo {author}
  {\bibfnamefont {M.}~\bibnamefont {Neeley}}, \bibinfo {author} {\bibfnamefont
  {A.~D.}\ \bibnamefont {O'Connell}}, \bibinfo {author} {\bibfnamefont
  {D.}~\bibnamefont {Sank}}, \bibinfo {author} {\bibfnamefont {M.}~\bibnamefont
  {Weides}}, \bibinfo {author} {\bibfnamefont {J.}~\bibnamefont {Wenner}},
  \bibinfo {author} {\bibfnamefont {T.}~\bibnamefont {Yamamoto}}, \bibinfo
  {author} {\bibfnamefont {Y.}~\bibnamefont {Yin}}, \bibinfo {author}
  {\bibfnamefont {J.}~\bibnamefont {Zhao}}, \bibinfo {author} {\bibfnamefont
  {A.~N.}\ \bibnamefont {Cleland}}, \ and\ \bibinfo {author} {\bibfnamefont
  {J.~M.}\ \bibnamefont {Martinis}},\ }\href@noop {} {\bibfield  {journal}
  {\bibinfo  {journal} {Phys. Rev. B}\ }\textbf {\bibinfo {volume} {84}},\
  \bibinfo {pages} {024501} (\bibinfo {year} {2011})}\BibitemShut {NoStop}%
\bibitem [{\citenamefont {Zazunov}\ \emph {et~al.}(2005)\citenamefont
  {Zazunov}, \citenamefont {Shumeiko}, \citenamefont {Wendin},\ and\
  \citenamefont {Bratus'}}]{ShumeikoPhonon}%
  \BibitemOpen
  \bibfield  {author} {\bibinfo {author} {\bibfnamefont {A.}~\bibnamefont
  {Zazunov}}, \bibinfo {author} {\bibfnamefont {V.~S.}\ \bibnamefont
  {Shumeiko}}, \bibinfo {author} {\bibfnamefont {G.}~\bibnamefont {Wendin}}, \
  and\ \bibinfo {author} {\bibfnamefont {E.~N.}\ \bibnamefont {Bratus'}},\
  }\href@noop {} {\bibfield  {journal} {\bibinfo  {journal} {Phys. Rev. B}\
  }\textbf {\bibinfo {volume} {71}},\ \bibinfo {pages} {214505} (\bibinfo
  {year} {2005})}\BibitemShut {NoStop}%
\bibitem [{\citenamefont {Vijay}\ \emph {et~al.}(2009)\citenamefont {Vijay},
  \citenamefont {Sau}, \citenamefont {Cohen},\ and\ \citenamefont
  {Siddiqi}}]{Vijay2009}%
  \BibitemOpen
  \bibfield  {author} {\bibinfo {author} {\bibfnamefont {R.}~\bibnamefont
  {Vijay}}, \bibinfo {author} {\bibfnamefont {J.~D.}\ \bibnamefont {Sau}},
  \bibinfo {author} {\bibfnamefont {M.~L.}\ \bibnamefont {Cohen}}, \ and\
  \bibinfo {author} {\bibfnamefont {I.}~\bibnamefont {Siddiqi}},\ }\href
  {\doibase 10.1103/PhysRevLett.103.087003} {\bibfield  {journal} {\bibinfo
  {journal} {Phys. Rev. Lett.}\ }\textbf {\bibinfo {volume} {103}},\ \bibinfo
  {pages} {087003} (\bibinfo {year} {2009})}\BibitemShut {NoStop}%
\bibitem [{\citenamefont {Vijay}\ \emph {et~al.}(2010)\citenamefont {Vijay},
  \citenamefont {Levenson-Falk}, \citenamefont {Slichter},\ and\ \citenamefont
  {Siddiqi}}]{Vijay2010}%
  \BibitemOpen
  \bibfield  {author} {\bibinfo {author} {\bibfnamefont {R.}~\bibnamefont
  {Vijay}}, \bibinfo {author} {\bibfnamefont {E.~M.}\ \bibnamefont
  {Levenson-Falk}}, \bibinfo {author} {\bibfnamefont {D.~H.}\ \bibnamefont
  {Slichter}}, \ and\ \bibinfo {author} {\bibfnamefont {I.}~\bibnamefont
  {Siddiqi}},\ }\href {http://arxiv.org/abs/1005.1110} {\bibfield  {journal}
  {\bibinfo  {journal} {Applied Physics Letters}\ }\textbf {\bibinfo {volume}
  {96}},\ \bibinfo {pages} {3} (\bibinfo {year} {2010})}\BibitemShut {NoStop}%
\end{thebibliography}%


%merlin.mbs apsrev4-1.bst 2010-07-25 4.21a (PWD, AO, DPC) hacked
%Control: key (0)
%Control: author (8) initials jnrlst
%Control: editor formatted (1) identically to author
%Control: production of article title (-1) disabled
%Control: page (0) single
%Control: year (1) truncated
%Control: production of eprint (0) enabled
\begin{thebibliography}{4}%
\makeatletter
\providecommand \@ifxundefined [1]{%
 \@ifx{#1\undefined}
}%
\providecommand \@ifnum [1]{%
 \ifnum #1\expandafter \@firstoftwo
 \else \expandafter \@secondoftwo
 \fi
}%
\providecommand \@ifx [1]{%
 \ifx #1\expandafter \@firstoftwo
 \else \expandafter \@secondoftwo
 \fi
}%
\providecommand \natexlab [1]{#1}%
\providecommand \enquote  [1]{``#1''}%
\providecommand \bibnamefont  [1]{#1}%
\providecommand \bibfnamefont [1]{#1}%
\providecommand \citenamefont [1]{#1}%
\providecommand \href@noop [0]{\@secondoftwo}%
\providecommand \href [0]{\begingroup \@sanitize@url \@href}%
\providecommand \@href[1]{\@@startlink{#1}\@@href}%
\providecommand \@@href[1]{\endgroup#1\@@endlink}%
\providecommand \@sanitize@url [0]{\catcode `\\12\catcode `\$12\catcode
  `\&12\catcode `\#12\catcode `\^12\catcode `\_12\catcode `\%12\relax}%
\providecommand \@@startlink[1]{}%
\providecommand \@@endlink[0]{}%
\providecommand \url  [0]{\begingroup\@sanitize@url \@url }%
\providecommand \@url [1]{\endgroup\@href {#1}{\urlprefix }}%
\providecommand \urlprefix  [0]{URL }%
\providecommand \Eprint [0]{\href }%
\providecommand \doibase [0]{http://dx.doi.org/}%
\providecommand \selectlanguage [0]{\@gobble}%
\providecommand \bibinfo  [0]{\@secondoftwo}%
\providecommand \bibfield  [0]{\@secondoftwo}%
\providecommand \translation [1]{[#1]}%
\providecommand \BibitemOpen [0]{}%
\providecommand \bibitemStop [0]{}%
\providecommand \bibitemNoStop [0]{.\EOS\space}%
\providecommand \EOS [0]{\spacefactor3000\relax}%
\providecommand \BibitemShut  [1]{\csname bibitem#1\endcsname}%
\let\auto@bib@innerbib\@empty
%</preamble>
\bibitem [{\citenamefont {Kulik}\ and\ \citenamefont
  {Omelyanchuk}(1975)}]{KO1_cpr}%
  \BibitemOpen
  \bibfield  {author} {\bibinfo {author} {\bibfnamefont {I.~O.}\ \bibnamefont
  {Kulik}}\ and\ \bibinfo {author} {\bibfnamefont {A.~N.}\ \bibnamefont
  {Omelyanchuk}},\ }\href@noop {} {\bibfield  {journal} {\bibinfo  {journal}
  {JETP Lett.}\ }\textbf {\bibinfo {volume} {21}},\ \bibinfo {pages} {96}
  (\bibinfo {year} {1975})}\BibitemShut {NoStop}%
\bibitem [{\citenamefont {Wilson}\ and\ \citenamefont
  {Prober}(2004)}]{Prober2004}%
  \BibitemOpen
  \bibfield  {author} {\bibinfo {author} {\bibfnamefont {C.~M.}\ \bibnamefont
  {Wilson}}\ and\ \bibinfo {author} {\bibfnamefont {D.~E.}\ \bibnamefont
  {Prober}},\ }\href {\doibase 10.1103/PhysRevB.69.094524} {\bibfield
  {journal} {\bibinfo  {journal} {Phys. Rev. B}\ }\textbf {\bibinfo {volume}
  {69}},\ \bibinfo {pages} {094524} (\bibinfo {year} {2004})}\BibitemShut
  {NoStop}%
\bibitem [{\citenamefont {Kaplan}\ \emph {et~al.}(1976)\citenamefont {Kaplan},
  \citenamefont {Chi}, \citenamefont {Langenberg}, \citenamefont {Chang},
  \citenamefont {Jafarey},\ and\ \citenamefont {Scalapino}}]{Scalapino1976}%
  \BibitemOpen
  \bibfield  {author} {\bibinfo {author} {\bibfnamefont {S.~B.}\ \bibnamefont
  {Kaplan}}, \bibinfo {author} {\bibfnamefont {C.~C.}\ \bibnamefont {Chi}},
  \bibinfo {author} {\bibfnamefont {D.~N.}\ \bibnamefont {Langenberg}},
  \bibinfo {author} {\bibfnamefont {J.~J.}\ \bibnamefont {Chang}}, \bibinfo
  {author} {\bibfnamefont {S.}~\bibnamefont {Jafarey}}, \ and\ \bibinfo
  {author} {\bibfnamefont {D.~J.}\ \bibnamefont {Scalapino}},\ }\href {\doibase
  10.1103/PhysRevB.14.4854} {\bibfield  {journal} {\bibinfo  {journal} {Phys.
  Rev. B}\ }\textbf {\bibinfo {volume} {14}},\ \bibinfo {pages} {4854}
  (\bibinfo {year} {1976})}\BibitemShut {NoStop}%
\bibitem [{\citenamefont {Kos}\ \emph {et~al.}(2013)\citenamefont {Kos},
  \citenamefont {Nigg},\ and\ \citenamefont {Glazman}}]{Kos2013}%
  \BibitemOpen
  \bibfield  {author} {\bibinfo {author} {\bibfnamefont {F.}~\bibnamefont
  {Kos}}, \bibinfo {author} {\bibfnamefont {S.~E.}\ \bibnamefont {Nigg}}, \
  and\ \bibinfo {author} {\bibfnamefont {L.~I.}\ \bibnamefont {Glazman}},\
  }\href {\doibase 10.1103/PhysRevB.87.174521} {\bibfield  {journal} {\bibinfo
  {journal} {Phys. Rev. B}\ }\textbf {\bibinfo {volume} {87}},\ \bibinfo
  {pages} {174521} (\bibinfo {year} {2013})}\BibitemShut {NoStop}%
\end{thebibliography}%

\end{document}

\title{Supplement to ``Single quasiparticle trapping in aluminum nanobridge Josephson junctions"}
%\author{E. M. Levenson-Falk} 
%\affiliation{Quantum Nanoelectronics Laboratory, Department of Physics, University of California, Berkeley CA 94720}
%\email{elevens@berkeley.edu}
%\author{F. Kos}
%\affiliation{Department of Physics, Yale University, New Haven CT 06520}
%\author{R. Vijay}
%\affiliation{Quantum Nanoelectronics Laboratory, Department of Physics, University of California, Berkeley CA 94720}
%\affiliation{Tata Institute for Fundamental Research, Mumbai, India}
%\author{L. Glazman} 
%\affiliation{Department of Physics, Yale University, New Haven CT 06520}
%\author{I. Siddiqi}
%\affiliation{Quantum Nanoelectronics Laboratory, Department of Physics, University of California, Berkeley CA 94720}

\maketitle

\section{Fabrication Details}
The device is fabricated using conventional double-angle shadow-mask evaporation and a liftoff process.  First, a two-layer PMMA/copolymer resist stack is spun onto a silicon substrate.  The nanobridge resonator structure is then defined lithographically in an SEM running at 30 keV.  The SQUID pattern is a  750 nm wide, $2\times2 \ \mu$m square washer, with two 100 nm gaps spanned by the 25 nm wide nanobridges.  After development in a solution of 1 part methyl isobutyl ketone to 3 parts isopropanol at $-15^\circ$ C, the sample is put in an electron-beam evaporator with a base pressure of $\sim 10^{-8}$ Torr.  The evaporation is performed in two stages.  First, 8 nm of aluminum are evaporated at normal incidence, depositing metal on the substrate everywhere exposed by the SEM beam.  Next, the sample is tilted in-situ to an angle of $36^\circ$.  This causes the nanobridge region to be occluded, as there is no longer a line-of-sight from the substrate to the evaporation source through the upper resist layer due to the steep aspect ratio of the nanobridge resist window.  We evaporate 80 nm of aluminum in this configuration, defining banks which are $8+80\cos(36^\circ) = 73$ nm thick, connected by the 8 nm bridge.  Finally, the resist is lifted off by soaking in acetone, leaving only the resonator with variable-thickness bridge defined.

\section{Extracting $q_0$ and $L_J$}
We model the nanobridge SQUID as two identical nanobridge junctions in parallel.  The resonant frequency of the device is given by $\omega_0 = (C(L+L_J/2))^{-1/2}$, where $C$ is the total capacitance, $L_J$ is the inductance of a single nanobridge, and $L$ is the linear inductance of the circuit.  Defining the junction participation ratio $q$ by 
\begin{align}
q(\phi) &\equiv\frac{ L_J(\phi)/2}{L + L_J(\phi)/2}\\
q_0 &\equiv q(\phi=0) ,
\end {align}
 one can write 
\begin{equation}
\omega_0(\phi) = \omega_0(0)[1 + q_0\frac{L_J(\phi) - L_J(0)}{L_J(0)}]^{-1/2}\,.
\label{eq:omega0}
\end{equation}
In order to fit $q_0$, we assume a KO-1 current-phase relation (CPR) for the nanobridges\cite{KO1_cpr}, thus giving
\begin{equation}
\omega_0(\phi) = \omega_0(0)[1 + q_0\frac{\sin\frac{\delta}{2}\tanh^{-1}\sin\frac{\delta}{2}}{1 - \sin\frac{\delta}{2}\tanh^{-1}\sin\frac{\delta}{2}}]^{-1/2}\,.
\end{equation}
Here, we use $\delta = \pi \phi$.  Fitting the flux tuning of the resonant frequency with this equation gives $q_0 = 0.015$.  We performed EM simulations of the oscillator structure in Microwave Office using the AXIEM solver and found good agreement with experimentally measured quantities.  These simulations gave $L = 1.2$ nH, and so we find $L_J(0) = 2 L q_0 / (1-q_0) = 36$ pH.

\section{Theoretical Model}
We consider a weak link with $N_e$ channels, i.e. $N_e$ Andreev states. Let $n_i =0,1$ denote the number of quasiparticles (QPs) trapped in the $i$-th channel. For a given configuration $\{n_i \}$ of QPs, we can find the resonant frequency of the circuit, $\omega_0(\{ n_i\})$. We denote the frequency-dependent response function as $S(\omega,\omega_0(\{n_i \}))$, since it depends on both $\omega$ and resonant frequency, which is a function of QP configuration. In the experiment we measure the average value of the response function, $\bar S (\omega)$:
\begin{equation}
\bar S (\omega) = \sum_{\{ n_i \}} p(\{n_i\}) S(\omega, \omega_0(\{ n_i \}))\,. \label{barS}
\end{equation}
where $p(\{n_i\})$ is the probability to find the system in the configuration $\{ n_i \}$. We assume it can be written as a product $p(\{n_i\}) = \prod_i p_i(n_i)$. The average number of QPs is then $\bar n_{trap} = \sum_i p_i$, where $p_i=p_i(n_i=1)$. Assuming $p_i \ll 1$, the probability to have zero QPs is $e^{-\bar n_{trap}}$. The expression for $\bar S (\omega)$ can be written as a convolution:
\begin{equation}
\bar S (\omega) = \int d\Omega F(\Omega) S(\omega, \Omega)\,, \label{Sconv}
\end{equation}
where
\begin{equation}
F(\Omega) = \sum_{\{ n_i \}} p(\{ n_i \}) \delta(\Omega - \omega_0 (\{ n_i \}))\,. \label{Fnew}
\end{equation}
 We distinguish two different cases -- large average number of QPs ($\bar n_{trap} \gg 1$) and small average number of QPs ($\bar n_{trap} \lesssim 1$). In the case of large average number of QPs, we evaluate $F(\Omega)$ using the saddle point approximation as follows:
\begin{align}
F(\Omega) = \int\frac{d\alpha}{2 \pi} \sum_{ \{ n_i \} } p(\{ n_i \}) e^{i \alpha (\Omega - \omega_0(\{ n_i \}))} = \int\frac{d\alpha}{2 \pi} \sum_{\{ n_i \}} p(\{ n_i \}) \exp[i \alpha ( \Omega - \omega_0^{(0)} - \sum_i \frac{\partial \omega_0}{\partial n_i} n_i ) ]\,.
\end{align}
Here we have expanded $\omega_0(\{ n_i \})$ around $\omega_0^{(0)}$, corresponding to no QPs, since each  QP only slightly shifts the resonant frequency. Writing the probability $p(\{ n_i \})$ as a product of $p_i$'s we get:
\begin{align}
F(\Omega) = \int \frac{d\alpha}{2 \pi} \exp[ i\alpha (\Omega - \omega_0^{(0)}) + \sum_i \ln (1 - p_i  + p_i  e^{-i\alpha \partial \omega_0/\partial n_i} ) ]\,.
\end{align}
The saddle point approximation of this integral then gives a Gaussian for $F(\Omega)$:
\begin{align}
F(\Omega) \propto \exp\left[- \frac{(\Omega - \omega_0^{(0)} -\sum_i p_i \partial \omega_0/\partial n_i)^2}{2 \sum_i p_i (\partial \omega_0/\partial n_i)^2}\right] \,,\label{Flarge}
\end{align}
yielding the Gaussian referred to in the main text with center frequency $\omega_0 = \omega_0^{(0)} + \sum_i p_i \frac{\partial \omega_0}{\partial n_i}$ and width $\delta \omega_0^2 = \sum_i p_i (\frac{\partial \omega_0}{\partial n_i})^2 $.  For $S(\omega, \Omega)$ we assume a  Lorentzian lineshape:
\begin{align}
S(\omega, \Omega) = \frac{\Gamma/\pi}{(\omega- \Omega)^2 + \Gamma^2}\,, \label{Snew}
\end{align}
where the width $\Gamma$ is easily measured from the resonance at zero flux.  Thus, in the case of large number of QPs, the resonance is a convolved Gaussian-Lorentzian with its maximum shifted from the 0-QP peak. The 0-QP peak is suppressed by a factor $e^{-\bar n_{trap}}$ and thus is too small to be observed.  In the case of a diffusive nanobridge we can find the exact expression for the inductance $L_J$ with a general distribution of QPs. The channel transmittivities $\tau$ in a diffusive nanobridge are given by the Dorokhov distribution:
\begin{equation}
\rho(\tau) = \frac{\pi \hbar G}{2 e^2} \frac{1}{\tau \sqrt{1-\tau}} \,,
\end{equation}
where $G$ is the normal-state conductance of the nanobridge. The number of channels is not well defined for this distribution, but we can define the effective number of channels as 
\begin{equation}
N_e = \frac{\left(\int_0^1d\tau \rho(\tau) \tau\right)^2}{\int_0^1d\tau \rho(\tau) \tau^2} = 3\pi \hbar G/2e^2.
\end{equation}
 Assuming that the quasiparticles in the nanobridge are distributed according to a given distribution $f(E)$, the inductance is given by:
\begin{align}
1/L_J = \int_0^1 d\tau \rho(\tau) \frac{e^2 \Delta \tau}{\hbar^2} \frac{\cos \delta + \tau \sin^4\frac{\delta}{2}}{\left(1 - \tau\sin^2 \frac{\delta}{2} \right)^{3/2}}  [1-2f(E_A(\delta,\tau))]\,,
\label{eq:LJ}
\end{align}
where $E_A(\delta,\tau)$ is the Andreev energy of a channel with transmittivity $\tau$ (see Eq.~(1) in the main text). It is clear from Eq.~(\ref{eq:LJ}) that the presence of QPs changes the expectation value of the inductance. Similarly, fluctuations of the number or distribution of trapped QPs cause fluctuations of $L_J$. Since the resonant frequency depends on the inductance through Eq.~(\ref{eq:omega0}), QPs therefore cause a resonant frequency shift and broaden the resonance. Assuming $f(E)$ is small and given by the Gibbs distribution with some low temperature $k_B T\ll \Delta-\Delta|\cos\frac{\delta}{2}|$, we get the following expressions:
\begin{align}
\omega_0^{(0)} - \omega_0 =  \omega_0^{(0)}  g(\delta) \frac{\bar{n}_{trap}}{N_e}\,, &\qquad
\delta \omega_0 ^2 = \frac{1}{4} (\omega_0^{(0)})^2 g^2(\delta)\frac{\bar{n}_{trap}}{N_e^2}\,, \label{eq:shiftwidth} \\
g(\delta) \equiv \frac{3q(\delta)}{2} & \frac{|\cos \frac{\delta}{2}|}{1- \sin\frac{\delta}{2}\tanh^{-1}\sin\frac{\delta}{2}}\,.
\end{align}
All quantities appearing in the above expressions are measured except $\bar{n}_{trap}$. Therefore, we may fit the resonance to the convolved Gaussian-Lorentzian described above with $\bar{n}_{trap}$ as the only fitting parameter. However, we find out that such a procedure does not fit the data well. Thus, we take the frequency shift and the width as independent fitting parameters, and extract $\bar{n}_{trap}$ from the best fit value of each of the parameters using Eq.~(\ref{eq:shiftwidth}). As explained in the main text, we find that the value of $\bar{n}_{trap}$ extracted from the width of the resonance is about a factor of 3 greater than the value $\bar{n}_{trap}$ extracted from the shift of the resonance.

In the case of small number of QPs ($\bar n_{trap} \lesssim 1$) the main resonance peak comes from the 0-QP resonance, which is suppressed by a factor $e^{-\bar n_{trap}} \lesssim 1$. Additional contributions to the response function come from the 1-QP and 2-QP resonances, while the multi-QP contributions are further suppressed. We can thus keep only terms corresponding to 0, 1, and 2 QPs in the expression for $F(\Omega)$, Eq.~(\ref{Fnew}):
\begin{align}
F(\Omega) = e^{-\bar n_{trap}}[ \delta(\Omega - \omega_0^{(0)}) +  \sum_i p_i \delta(\Omega - \omega_0^{(i)}) +  \sum_{i,j} p_i p_j \delta(\Omega - \omega_0^{(i,j)})]\,, \label{Fsmall}
\end{align}
where $\omega_0^{(0)}$, $\omega_0^{(i)}$, and $\omega_0^{(i,j)}$ are the resonant frequencies at zero, one, and two trapped QPs, respectively; $i$ and $j$ label channels.

%Convolving Eq.~(\ref{Fsmall}) with the Lorentzian of Eq.~(\ref{Snew}) we get the expression for the resonant curve in the case of small number of QPs.
We write the resonant frequency with a QP trapped in a channel with transmittivity $\tau^{(i)}$ and energy $E_A^{(i)}$ as
\begin{align}
\omega_0^{(i)} \equiv \omega_0(\tau^{(i)}) = \omega_0^{(0)} - \frac{q}{2} \omega_0^{(0)} \cdot  L_J^{(0)}\Delta\frac{1}{L_J^{(i)}}\, , \quad \Delta\frac{1}{L_J^{(i)}} = \frac{2 e^2\tau_i\Delta\left( \cos\delta + \tau_i \sin^4\frac{\delta}{2} \right)}{\hbar^2(E_A^{(i)}/\Delta)^3}\,. \label{omegai}
\end{align}
Here $\tau_i$ and $\Delta\frac{1}{L_J^{(i)}}$ are the transmittivity of the $i$th channel and its contribution to the inductance, respectively.
%\begin{align}
%\omega_0^{\text{min}} = \omega_0^{(0)} - \frac{2q\omega_0^{(0)}}{N_e} \frac{|\cos\frac{\phi}{2}|}{1-\sin\frac{\phi}{2}\artanh\sin\frac{\phi}{2}}  
%\end{align}

 Using Eq.~(\ref{omegai}) in expression (\ref{Fsmall}) and transforming the sums into integrals by averaging over the distribution of transmission coefficients, we find
\begin{align}
F(\Omega) = e^{-\bar n_{trap}}[ \delta (\Omega -\omega_0^{(0)}) +  \int\limits_0^1 d\tau \rho(\tau) p(E_A(\tau)) \delta(\Omega - \omega_0(\tau)) \notag\\
+\int\limits_0^1 d\tau_1 \int\limits_0^1 d\tau_2 \rho(\tau_1)\rho(\tau_2 )p(E_A(\tau_1) )p(E_A(\tau_2)) \delta(\Omega - \omega(\tau_1,\tau_2))].
\label{Fsmall2}
\end{align}
 We assume a Gibbs distribution for the QPs:
\begin{align}
p(E_A(\tau)) \equiv \bar n_{trap} f(E_A(\tau)) = \bar n_{trap} \ \mathcal{N}e^{-E_A(\tau)/k_B T}\,,
\end{align}
where the normalization factor $\mathcal{N}$ is determined from the condition $\int d\tau \rho(\tau)p(E_A(\tau))=\bar n_{trap}$.  Finally, convolving the expression for $F(\Omega)$ with the Lorentzian lineshape given in Eq.~(\ref{Snew}), we find the response function (see Eq.~(2) in the main text):
\begin{align}
\bar S (\omega) = P_0 S(\omega, \omega_0^{(0)}) + P_1 \int\limits_0^1 d\tau \rho(\tau) f(E_A(\tau)) S(\omega, \omega_0(\tau)) \notag\\
+ P_2 \int\limits_0^1 d\tau_1\int\limits_0^1 d\tau_2 \ \rho(\tau_1)  \rho(\tau_2) f(E_A(\tau_1))f(E_A(\tau_2)) S(\omega, \omega_0(\tau_1,\tau_2))\,,
\label{Sfinal2}
\end{align}
where $P_k$ is the probability of having exactly $k$ trapped QPs.  Our data typically do not agree with the Poisson-distributed values of $P_k = \bar{n}_{trap}^k \exp(-\bar n_{trap})/k! $ that are expected for rare, independent trapping events. Because of this, we allow the $P_k$ to vary independently, improving the fits. We may also use temperature as an additional fitting parameter, but even allowing $T$ to vary freely gives $P_k$ which are not Poisson-distributed.

\section{Retrapping Time Measurements}
We probe the resonator at its 0-QP resonant frequency with a constant tone.  We mix the reflected tone down to dc with an IQ demodulation setup and measure the two quadratures with a fast digitizer circuit.  We then pulse the QP excitation tone at a frequency which is significantly above $\Delta_A$.  This removes some percentage of the QPs out of the junction trap states. We test the pulse's efficacy by increasing its power, and determine that at the power at which the response saturates there are no trapped QPs.  Fig. S1(a) shows a sample trace at 10 mK monitoring one quadrature of the measurement as the pulse is turned on and off, averaged over many iterations.  The resonance shifts to a new steady state with the pulse on and decays back to its old value after it is turned off, with an exponential envelope for both processes.  The first exponential corresponds to the clearing of QPs from the junction.  It has a time constant which depends linearly on the excitation pulse power, and is roughly independent of $\phi$ as shown in Fig. S1(b).  The second exponential corresponds to the retrapping of QPs as the system returns to steady state.  It has a time constant which is independent of excitation pulse amplitude and duration, and which decreases as a function of $\phi$ from $60$ to $20 \ \mu$s as $\phi$ increases from 0.35 to 0.55; see Fig. S1(b).  Below $\phi=0.35$ the pulse response signal was too small to accurately measure.

\begin{figure}
\includegraphics{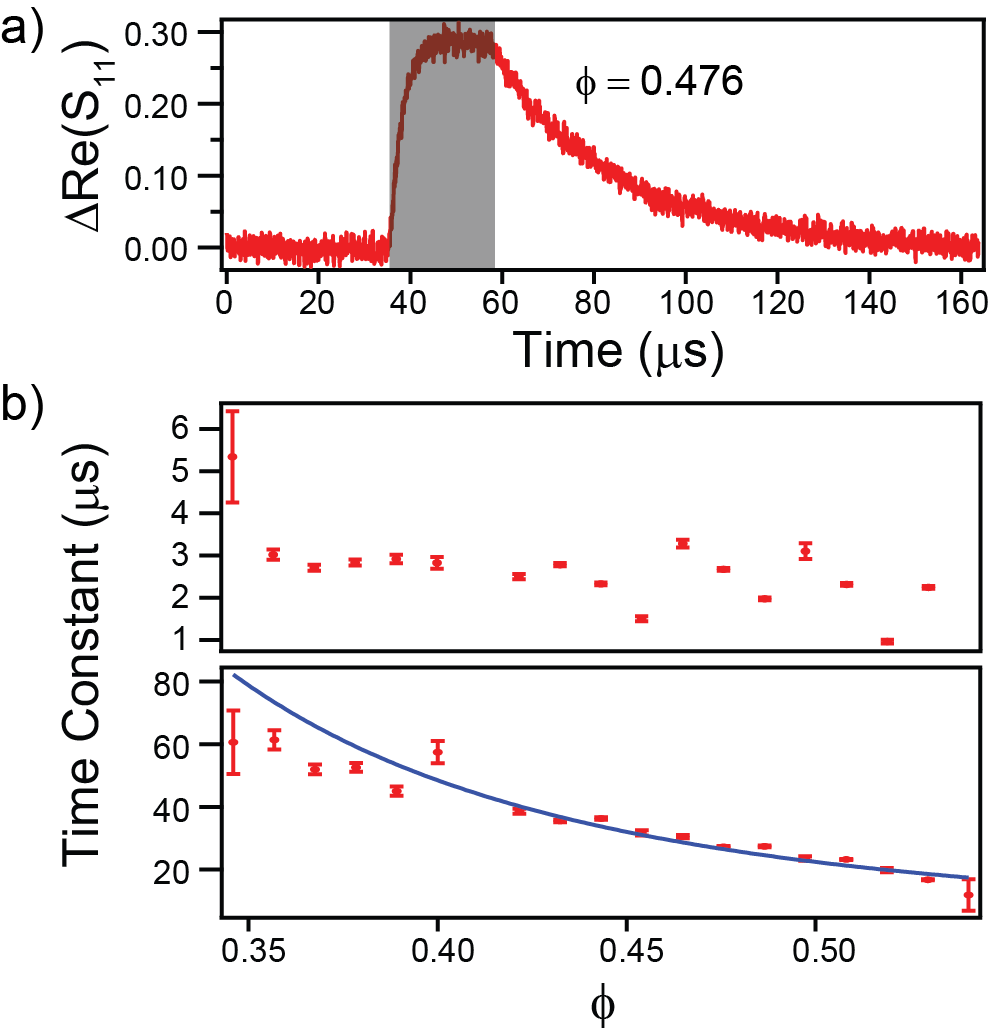}
\caption{\label{figs1} (Color online) (a) Representative measurement of QP excitation and retrapping at $\phi = 0.476$.  The shaded region indicates the time during which the excitation pulse is on.  We fit the exponential rise and fall to extract excitation and retrapping time constants, respectively.  (b) QP excitation (top) and retrapping (bottom) time constants as a function of flux.  The excitation time is roughly constant as a function of flux for a given power, while the retrapping time decreases with $\phi$, in qualitative agreement with the theoretical prediction of electron-phonon relaxation (solid blue line).}
\end{figure}

We attempt to fit the retrapping time by assuming electron-phonon relaxation as the dominant mechanism. The crude model we use disregards the influence of disorder on the electron-phonon interaction, and the effect of bridge geometry. The electron-phonon interaction Hamiltonian then takes the form:
\begin{equation}
H_{e-ph} = \frac{1}{\sqrt{N}}\sum_{\bf{k},\bf{q},\sigma}\alpha q^{1/2}(a^\dag_{\bf{k}+\bf{q},\sigma}a_{\bf{k},\sigma}b_{\bf{q}} + a^\dag_{\bf{k},\sigma}a_{\bf{k}+\bf{q},\sigma}b^\dag_{\bf{q}})\,,
\end{equation}
where $a$ and $b$ are the annihilation operators and $\bf{k}$ and $\bf{q}$ are the momenta for electrons and phonons, respectively. We want to relate the retrapping time in our experiment, $\tau_T$, to the QP recombination time $\tau_R$ in a bulk superconductor that was measured in earlier experiments~\cite{Prober2004}. The recombination rate of a QP with momentum $\mathbf{p}$ in the superconductor is given by Fermi's Golden Rule as:
\begin{align}
w_R(\mathbf{p}) &= \frac{2\pi}{N}\sum_{\mathbf{p_2}}\sum_{\mathbf{q}}\alpha^2 q |u_p v_{p_2} + u_{p_2}v_p|^2 (N_q+1) f_{p_2} \delta_{\mathbf{p}+\mathbf{p_2},\mathbf{q}} \delta(E_p + E_{p_2}-\omega_q)\,,
\end{align}
where $u_p$ and $v_p$ are Bogoliubov amplitudes of a QP with momentum $\mathbf{p}$ and energy $E_p$, and $N_q$, $f_p$ are distribution functions of phonons and QPs, respectively. In the case where the QP energy is close to the gap, $E_p \approx \Delta$, and  the QPs are Gibbs-distributed at a low temperature $k_B T \ll \Delta$, the integrals in the above expression for the recombination rate can be evaluated to give\cite{Scalapino1976}
\begin{equation}
w_R = 4\pi x_{qp}b\Delta^3 \,,
\label{eq:wR}
\end{equation}
 where $b = 2V\alpha^2/\pi^2Ns^3v_F$, with $s$ and $v_F$ representing the speed of sound and Fermi velocity, respectively. The recombination time is defined as $\tau_R = 1/w_R$.

Next we consider the single-channel point contact with one Andreev state $\alpha_{A\sigma}^\dagger|0\rangle$, which can be expressed in terms of the electron operators as:
\begin{equation}
\alpha_{A\sigma} = \sum_{\mathbf{k}} \left( u_{A\mathbf{k}}^{L*} a_{\mathbf{k}\sigma}^{L} + v_{A\mathbf{k}}^{L*} a_{\mathbf{k}-\sigma}^{L\dagger} + u_{A\mathbf{k}}^{R*} a_{\mathbf{k}\sigma}^{R} + v_{A\mathbf{k}}^{R*} a_{\mathbf{k}-\sigma}^{R\dagger}\right) \,.
\end{equation}
Indices $L$ and $R$ denote left and right lead, respectively. The retrapping rate $w_T$ can be written as $w_T = \sum_\mu w_{\mu \to A} f(E_\mu)$, where the sum goes over all single-QP states and $f(E_\mu)$ is the initial distribution function of QPs. The transition rate from a given QP state $\mu$ to the Andreev level is given by Fermi's Golden Rule as:
\begin{align}
w_{\mu\to A} = 2\pi  \sum_{\mathbf{q}} | \langle 0 | \alpha_{A\sigma_A}b_q H_{e-ph} \alpha_{\mu \sigma}^\dagger |0\rangle |^2 \delta (E_\mu - E_A -\omega_q)\,.
\end{align}
The retrapping rate can then be expressed in the form:
\begin{align}
w_T = \frac{-i}{N} &\sum_{\mathbf{q}}  \alpha^2 q f(E_A+\omega_q) \sum_{i,j} \sum_{\mathbf{k}_1,\mathbf{k}_4} \big[  u^{i*}_{A\mathbf{k}_1} u^j_{A\mathbf{k}_4} \mathcal{G}_{ij}(\mathbf{k}_1-\mathbf{q},\mathbf{k}_4-\mathbf{q};E_A+\omega_q) \notag\\
 +&  v^{i*}_{A\mathbf{k}_1} v^j_{A\mathbf{k}_4} \mathcal{G}_{ij}(\mathbf{k}_4 + \mathbf{q}, \mathbf{k}_1 + \mathbf{q}; -E_A-\omega_q) + v^{i*}_{A\mathbf{k}_1} u^j_{A\mathbf{k}_4} \mathcal{F}^+_{ij}(\mathbf{k}_1+\mathbf{q},\mathbf{k}_4-\mathbf{q};E_A+\omega_q) \notag\\
+& u^{i*}_{A\mathbf{k}_1} v^j_{A\mathbf{k}_4} \mathcal{F}_{ij}(\mathbf{k}_1-\mathbf{q},\mathbf{k}_4+\mathbf{q};E_A+\omega_q)  \big]\,.
\label{eq:wA}
\end{align}
Here $\mathcal{G}_{ij} =G_{ij}^R - G_{ij}^A$ and $\mathcal{F^\pm}_{ij} =F_{ij}^{\pm R} - F_{ij}^{\pm A}$ are electron-electron Green's functions in the point contact. The indices $i, j$ in the above sum are going over the values $L, R$. Describing the point contact with transmittivity $\tau$ using the tunneling Hamiltonian\cite{Kos2013}, it is possible to find the exact expressions for the Green's functions and the Bogoliubov amplitudes of the Andreev states. However, the integrals in Eq.~(\ref{eq:wA}) are difficult to evaluate for arbitrary $\tau$, so we only consider the limit of a shallow Andreev level, in which case the amplitudes are given by:
\begin{align}
|u^{L,R}_{A\mathbf{k}}|^2 = \frac{(\Delta^2-E_A^2)^{3/2}}{2\pi\nu_0(E_k^2-\Delta^2)^2} \,,\\
v^L_{A\mathbf{k}} = e^{i\delta/2}u^L_{A\mathbf{k}} = e^{-i\delta/2}u^{R*}_{A\mathbf{k}}=v^{R*}_{A\mathbf{k}}\,,
\end{align}
where $\nu_0$ is the normal-state density of states at Fermi level. For the crude estimate of the retrapping time, we use the bulk superconductor Green's functions. Taking as the initial distribution of QPs a Gibbs distribution at a low temperature $k_BT \ll \Delta-E_A$, we can evaluate Eq.~(\ref{eq:wA}):
\begin{equation}
w_T = \pi x_{qp} b \Delta (\Delta - E_A)^2\,.
\end{equation}
Defining the retrapping time as $\tau_T = 1/w_T$ and comparing with Eq.~(\ref{eq:wR}), we get:
\begin{equation}
\tau_T = \left(  \frac{2\Delta}{\Delta-E_A} \right)^2 \tau_R \label{retrapping}\,.
\end{equation}
A fit to the relaxation time measurements using this functional form, with $E_A(\phi) =\Delta |\cos\frac{\phi \pi}{2}| $, is shown in Fig. S1(b). The functional form derived for $\tau_T(\phi)$ agrees reasonably with the measurements. However, the value $\tau_R=0.3\mu$s which we extract with the help of Eq.~(\ref{retrapping}) is much lower than the value $\tau_R = 100\mu$s reported in earlier experiments at 250mK\cite{Prober2004}. We note, however, that our derivation of Eq.~(\ref{retrapping}) was very crude; we assumed the clean superconductor limit, did not take into account the geometry of the junction, and considered only the limit of a shallow Andreev level. A more complete theoretical treatment may be able to resolve the discrepancy.

\bibliography{supplementv2}